\documentclass[twocolumn,showpacs,preprintnumbers,amsmath,amssymb]{revtex4}
\usepackage{dcolumn}
\usepackage{epsfig}

\newcommand{\be}{\begin{equation}}
\newcommand{\ee}{\end{equation}}
\newcommand{\bea}{\begin{eqnarray}}
\newcommand{\eea}{\end{eqnarray}}
\newcommand{\ww}{{w}}

\begin{document}

\title{Resonant excitations of the 't~Hooft-Polyakov monopole}
\author{P\'eter Forg\'acs and Mikhail S. Volkov}
\affiliation{Laboratoire de Math\'ematiques et Physique Th\'eorique, CNRS-UMR 6083,
Universit\'e de Tours, Parc de Grandmont,
37200 Tours, FRANCE}

\begin{abstract}
 {The spherically symmetric magnetic monopole in an SU(2) gauge theory coupled to
 a massless Higgs field is shown to possess an infinite number of resonances or
 quasinormal modes.
These modes are eigenfunctions of the isospin 1 perturbation equations
with complex eigenvalues,  $E_n=\omega_n-i\gamma_n$, 
satisfying the outgoing radiation condition. 
For $n\to\infty$, their frequencies $\omega_n$
approach the mass of the vector boson, $M_W$,
while their lifetimes $1/\gamma_n$ tend to infinity.
The response of the monopole to an arbitrary initial perturbation is largely
determined by these resonant modes, whose collective effect leads to the  formation of
a long living breather-like excitation characterized by
pulsations with a frequency approaching $M_W$ and with an amplitude decaying 
at late times as $t^{-5/6}$. }
\end{abstract}
\maketitle

Magnetic monopoles -- magnetically charged finite energy solutions in field theories with
spontaneously broken gauge symmetries \cite{mon} -- play an important role in a number
of field theoretic considerations.
They can account for the quantization of the electric charge, catalyze the decay of proton,
might be among the relevant topological defects in the universe,
determine supersymmetric vacua, etc. --
see \cite{monrev} for reviews.

In the present Letter we point out yet another interesting aspect of the monopole
which seems to have gone unnoticed so far -- the existence of its quasinormal modes (QNM)
or resonant excitations. We show  that the
Bogomol'nyi-Prasad-Sommerfield (BPS) monopole admits an infinite number of QNM.
They manifest themselves as resonance peaks in the low energy
scattering cross section of
isospin $1$ scalar particles. 
The QNM can also be described as {\sl complex energy} solutions
of the isospin $1$ small fluctuation equations  that are
regular at the origin and satisfy the {\sl outgoing radiation} condition
at spatial infinity.  For previous studies of
the {\sl real  energy} small fluctuations around the
 monopole we refer to \cite{bais}.

We shall demonstrate, in particular, that the QNM of the monopole lead to a {\sl universal
late time behavior} of the perturbed monopole by giving rise collectively to
a quasiperiodic, long living excitation whose amplitude decays as $t^{-5/6}$
at late times ($t$ being the standard Minkowski time).
In a recent numerical study of Fodor and R\'acz a breather-like excitation
of the monopole
retaining a considerable fraction of the energy of the external perturbation
has actually been observed \cite{Racz}. It has been one of our aims
to explain their observations. 

It is worth noting the striking analogy of 
the $t^{-5/6}$ asymptotic behavior of the monopole  
with the late time evolution of massive fields in
black hole spacetimes \cite{5/3}. Black holes also possess 
the QNM, which  have been 
 much studied recently (see e.g. \cite{BH}).

We consider a Yang-Mills-Higgs (YMH) theory
with gauge group SU(2) defined by the Lagrangian
\be                                 \label{1}
{\cal L}=-\frac14\,F^a_{\mu\nu}F^{a\mu\nu}+\frac12
D_\mu\Phi^a D^{\mu}\Phi^a-
\frac{\lambda}{4}(\Phi^a\Phi^a-1)^2.
\ee
Here
$ F^a_{\mu\nu}=\partial_{\mu} A^a_{\nu} - \partial_{\nu} A^a_{\mu}
+ \varepsilon_{abc}A^b_{\mu} A^c_{\nu}$ is the non-abelian field strength tensor,
$D_\mu\Phi^a=\partial_\mu\Phi^a+\varepsilon_{abc}A^b_\mu\Phi^c$
denotes the covariant derivative, and in our units
the mass of the gauge bosons is equal to one, $M_W=1$, while the mass of the Higgs
particle is $\sqrt{2\lambda}\,$.
The energy-momentum tensor of the theory defined by Eq.(\ref{1}) is
$T^\mu_\nu=-F^a_{\nu\rho}F^{a\mu\rho}
+D^\mu\Phi^a D_\nu\Phi^a
-\delta^\mu_\nu {\cal L}$.

We restrict our analysis to the `minimal' spherically symmetric
sector, where the ansatz for the YMH fields is given by $A^a_0=0$,
\be                                     \label{3}
A^a_i=\varepsilon_{aik}\frac{x^k}{r^2}(1-W(t,r))\,,
~~\Phi^a=\frac{x^a}{r^2}\,H(t,r)\,,
\ee
where $a,i,k=1,2,3$ and $r^2=x^kx^k$\,.
With $\Box\equiv
\frac{\partial^2}{\partial t^2 }-\frac{\partial^2}{\partial r^2 }$
the YMH equations reduce to
\bea                 \label{5}
(r^2\Box
+{W}^2+{H}^2-1){W}=0\,,  \nonumber \\
(r^2\Box+
2{W}^2+\lambda ({H}^2-r^2)){H}=0\,.
\eea
The static, finite energy solution of these equations
is the 't~Hooft-Polyakov monopole \cite{mon}.
For the special case, $\lambda=0$, the solution is analytically known -- the BPS monopole, 
\be                               \label{6}
W(r)=\frac{r}{\sinh r}\,,\quad
H(r)=r\coth r-1\,. 
\ee

We shall consider small fluctuations around the static monopole background:
$W\to W(r)+w(t,r)$ and $H\to H(r)+\sqrt{2}\,h(t,r)$.
Linearizing Eqs.(\ref{5}) with respect to $w$ and $h$,  we obtain
\bea
(r^2\Box+3W^2+H^2-1)w=-q\,{2\sqrt{2}WH}\,h,&& \label{2:1} \\
(r^2\Box+{2W^2}+\lambda(3H^2-r^2))h=-
q\,2\sqrt{2}WH\,w.&&   \label{2:2}
\eea
Here an auxiliary parameter, $q$, has been introduced,
presently $q=1$.
In this Letter we concentrate on the case $\lambda=0$,
when $W,H$ are given by Eq.(\ref{6}).

 {\bf Resonant scattering.--}
First we shall demonstrate that Eqs.(\ref{2:1}),(\ref{2:2}) describe
resonance phenomena indeed.
Separating the variables as
$w=\Re(e^{-i \omega t}w_\omega(r))$ and
$h=\Re(e^{-i \omega t}h_\omega(r))$ with real $\omega$,
Eqs.(\ref{2:1}),(\ref{2:2}) become a standard
two-channel Schr\"odinger system. As it will be clear from what follows, the
frequency spectrum is continuous, $\omega^2\geq~0$.
Regular solutions of Eqs.(\ref{2:1}),(\ref{2:2}) have to satisfy the conditions 
$w_\omega\sim h_\omega\sim r^2$ for $r\to 0$,
and they can be normalized for $r\to\infty$ such that
\be                                                              \label{stat}
h_\omega(r)\to \sin(\omega r+\delta(\omega))\,,\quad
w_\omega(r)\to C(\omega)  r^{\frac{1}{\nu}} e^{-\nu r}\,,
\ee
where  $\nu=\sqrt{1-\omega^2}$. The $h$-field is massless,
so it oscillates as $r\to\infty$ for any value of $\omega$.
The $w$-field is massive,
and for $\omega^2<1$ it shows a bound state type behavior,
with exponential decay as $r\to\infty$.
The fact that the $w$-field is {\it non-radiative} for $\omega^2<1$ plays the
crucial role in our analysis, and below we shall concentrate
to this frequency range.
Eqs.(\ref{2:1}),(\ref{2:2}) describe in this case the scattering of a massless
$h$-radiation on the monopole surrounded by a confined
massive $w$-field. We note that for $\omega^2<1$ this is effectively a {\it one-channel}
scattering problem, so that the scattering cross section is given by  the standard formula
$\sigma(\omega)=({4\pi}/{\omega^2})\sin^2(\delta(\omega))$.

\begin{figure}
\epsfxsize=9cm
\epsfysize=4cm
\centerline{\epsffile{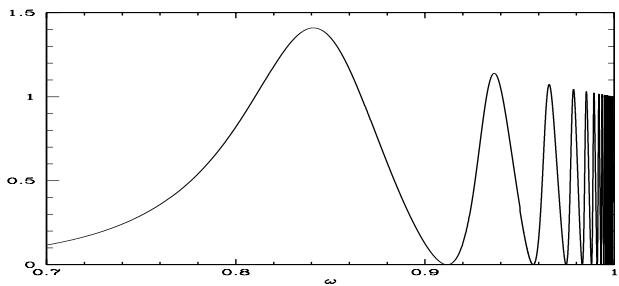}}
\caption{The scattering cross section $\sigma(\omega)$.
}
\label{fig2}
\end{figure}

It is worth noting that the interaction of the $h$-field with the
monopole is  in fact {\it short range}, so that the cross
section is finite. We integrate Eqs.(\ref{2:1}),(\ref{2:2})
numerically to obtain $w_\omega(r)$ and $h_\omega(r)$ subject to
the boundary conditions (\ref{stat}).
The resulting cross section $\sigma(\omega)$
shown in Fig.\ref{fig2} exhibits a sequence of resonant peaks
accumulating near the value $\omega=1$. This can be so interpreted that for
certain energies of the incident $h$-radiation the monopole core
gets strongly excited.

\begin{figure}
\epsfxsize=9cm
\epsfysize=4cm
\centerline{\epsffile{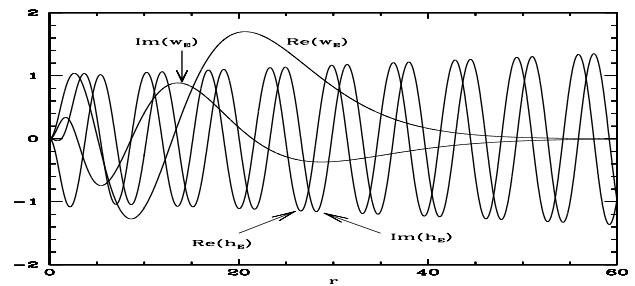}}
\caption{The complex solutions of Eqs.(\ref{2:1}),(\ref{2:2}) for $n=3$.
}
\label{fig3}
\end{figure}

{\bf QNM -- numerical results.--} The scattering resonances can
be usually related to the quasinormal modes --
complex energy solutions satisfying the purely outgoing wave condition at $r=\infty$.
To construct the QNM, we integrate Eqs.(\ref{2:1}),(\ref{2:2})
with $w=\Re(e^{-i E t}w_E(r))$ and $h=\Re(e^{-i Et}h_E(r))$, where
 $w_E$ and $h_E$ are complex, the energy $E=\omega-i\gamma$, and
\be \label{compl}
 Ar^2\leftarrow h_E(r)\to e^{iEr},~~~ Br^2\leftarrow w_E(r)\to C r^{\frac{1}{\nu}}
e^{-\nu r}
\ee
for $0\leftarrow r\to\infty$.
With the `shooting to a fitting point' numerical method we find a discrete
family of global solutions $w_E(r)$, $h_E(r)$ subject to the boundary conditions (\ref{compl})
labeled by $n=1,2,\ldots $,  the
number of nodes of $\Im(w_E(r))$ (see Fig.\ref{fig3}).
Notice that $h_E\sim e^{i\omega r+\gamma r}$
{\sl grows} at infinity -- the QNM are not  physical solutions themselves,
but only approximate such solutions for a fixed $r$ and for $t\to\infty$.
The first
$10$ eigenvalues $E_n=\omega_n-i\gamma_n$ are listed in Table~I.

Table I clearly indicates that $\omega_n\to 1$ and $\gamma_n\to 0$ for growing $n$.
It seems that the QNM can be obtained for any $n$, thus
comprising an infinite family. The values of
$\omega_n$ coincide well with the positions of the
resonance peaks shown in Fig.\ref{fig2}.

{\bf QNM -- qualitative analysis.--}\
The existence of the QNM
of the BPS monopole can be qualitatively understood as follows.
Let us consider
$q$ introduced in Eqs.(\ref{2:1}),(\ref{2:2})
as a free parameter, $q\in[0,1]$, and
denote the corresponding solutions by
$w_{(q)}$ and $h_{(q)}$.
For $q=0$ the equations decouple.
Setting $h_{(0)}(t,r)= e^{- i\omega t}(C_{+}h_{+}+C_{-}h_{-})$
with constant $C_{\pm}$, Eq.(\ref{2:2})
is then solved by
\be\label{h}
h_{\pm}(r)=(\coth r\mp i\omega)e^{\pm i\omega r}.
\ee
Eq.(\ref{2:1}) with $w_{(0)}(t,r)= e^{-i\omega t}w(r)$
reduces to the eigenvalue problem
\be \label{bound}
\left(-
\frac{d^2}{d r^2}+\frac{3W^2+H^2-1}{r^2}\right)w(r)=\omega^2 w(r).
\ee
The potential  in this equation has an attractive Coulombian tail,
since it behaves as
$1-2/r+O(e^{-r})$ for $r\to\infty$, thus there are
infinitely many bound states, $w(r)=w_n(r)$
for $\omega^2=\omega^2_n$, $n=1,2,\ldots $.
Several low lying $\omega_n$'s are given in Table II.
The $n$-th eigenfunction
$w_n(r)$  has $n-1$ nodes in the interval $r\in[0,\infty)$ and
can be normalized by the condition $\int_0^\infty w_n^2dr=1$.
For large $n$ the $w_n(r)$'s extend to the
asymptotic region where the potential is Coulombian.
As a result, they can be well approximated by solutions of the
hydrogen atom problem. This implies that for $n\to\infty$ one has
$\omega_n^2=1-1/n^2+O(n^{-3})$ and also that
$w_n\sim n^{-3/2}$. More precisely,
for $\omega=1$
Eq.(\ref{bound}) admits a limiting solution
with infinitely many nodes, $w_\infty(r)$, which itself is not normalizable,
but $w_\infty(r)$ corresponds to the pointwise limit of $w_n(r)$; i.e.
for a {\sl fixed} $\bar{r}$
\be                       \label{large}
\lim_{n\to\infty}w_n(r)=n^{-3/2}w_\infty(r)~~~~~~~~~\forall r<\bar{r}\,.
\ee

Let us consider a solution of Eqs.(\ref{2:1}),(\ref{2:2})
given by $h_{(0)}=0$
and
$w_{(0)}= \Re(A_n e^{-i\omega_n t}w_{n}(r))$,
where $A_n$ is a constant.
``Switching on'' a small value of the coupling between the channels,
$q\ll 1$, the $w$-bound state
will start loosing its energy to the $h$-channel, where
this energy will be radiated to infinity.
To find approximatively the corresponding solutions $h_{(q)}$
and $w_{(q)}$ by successive iterations,
we solve first Eq.(\ref{2:2}) for $h_{(q)}$ by replacing $w$ by $w_{(0)}$
in its right hand side.
The solution is regular at the origin and reduces to an outgoing wave at infinity:
$h_{(q)}\sim qA_ne^{i\omega_n (r-t)}$ as $r\to\infty$.

Since there is now an outgoing flux of the $h$-radiation,
the energy of the $w$-bound state will be slowly decreasing.
In the adiabatic approximation this process is described by
a decrease of the amplitude
of the $\ww$-field by replacing $A_n\to A_n(t)$.
To determine $A_n(t)$
we use the law of energy
conservation, $\partial_\mu T^\mu_0=0$,
whose integral form is
\be\label{flux}
\frac{d}{dt}\left(\int_0^\infty r^2T^0_0 dr \right)=
-\lim_{r\to\infty}r^2 T^r_0\, .
\ee
Expanding $T^\mu_\nu$ up to terms
quadratic in $w$ and $h$, 
the expression on the left in Eq.(\ref{flux})  is proportional to $\frac{d}{dt}A_n^2$ and
determines the  decrease of the bound state energy.
The expression on the right is proportional to $A_n^2$ and gives  the
energy flux at infinity.
Thus
$\dot{A_n}=-\gamma_n A_n$, from which $A_n(t)=c_ne^{-\gamma_n t}$ with a
constant $c_n$. The coefficient $\gamma_n$ is given by
\be\label{gamma}
\gamma_n=q^2\frac{4}{\omega_n^2(1+\omega_n^2)}
{\left(\int_0^\infty \frac{h_{+}-h_{-}}{2ir^2}\,WH w_n dr \right)^2 }.
\ee
Summarizing, upon switching on a small value of $q$,
the stationary bound state of the $w$-field, $w_n(r)$, becomes quasistationary
and is approximately given by
\be\label{approx}
{w}_{(q)}= \Re(e^{-i E_n t}c_n w_n)\,,\quad
{h}_{(q)}= q\,\Re(e^{-i E_n t}c_n h_n)\,,
\ee
where $E_n=\omega_n-i\gamma_n$, and
$h_n(r)$ can be expressed by quadratures in terms of $w_n(r)$.
Note the appearance of a complex energy, $E_n$, in
Eq.(\ref{approx}).
Evaluating the integral in Eq.(\ref{gamma}) (see Table II)
shows that  $\gamma_n$ {decreases} as $n$ grows.
This feature
can be understood qualitatively by viewing the QNM as quasibound states of the
massive $w$-field slowly decaying via an energy transfer to the
$h$ channel. For large $n$ the coupling between the $w$ and $h$
channels becomes weaker and weaker, since it is proportional to
$WH w_n\sim n^{-3/2}$, and so the decay becomes more
and more unlikely. Using the asymptotic relation (\ref{large})
in Eq.(\ref{gamma}) yields $\gamma_n\sim n^{-3}$ for $n\to\infty$.
\begin{table}
\caption{The eigenvalues of the first ten QNM}
\begin{ruledtabular}
\begin{tabular}{cccc} 
n &  $\omega_n-i\gamma_n$ & $n$  &  $\omega_n-i\gamma_n$ \\ \hline
1 &  $0.8473-i\, 0.5077\times 10^{-1}$  & 6 & $0.9888-i\, 0.8396\times 10^{-3}$  \\
2 & $0.9332-i\, 0.1384\times 10^{-1}$  &  7 & $0.9915-i\,0.5499\times 10^{-3}$  \\
3 & $0.9637-i\,0.5218\times 10^{-2}$  &  8 & $0.9933-i\,0.3794\times 10^{-3} $ \\
4 & $0.9774-i\,0.2488\times 10^{-2} $ &  9 & $0.9946-i\,0.2731\times 10^{-3}$  \\
 5 & $0.9846-i\, 0.1375\times 10^{-2}$  &10 & $0.9956-i\,0.2030\times 10^{-3} $ \\
\end{tabular}
\end{ruledtabular}
\end{table}

The approximative formulas derived above
under the assumption $q\ll 1$ give in fact, when straightforwardly extrapolated to $q=1$,
a good approximation for $\omega_n-i\gamma_n$ (see Tables I, II).
This approximation is actually getting better with growing $n$.
We can therefore use the asymptotic relations derived above
to obtain for $n\gg 1$
\be                                     \label{llarge}
1-\omega_n^2=n^{-2}+O(n^{-3})\,,\quad \gamma_n=bn^{-3}.
\ee
Using the results of Table I, $b\approx 0.2$.

\begin{table}
\label{tab1}
\caption{Values of $\omega_n$ and $\gamma_n$ defined by
Eqs.(\ref{bound}),(\ref{gamma})}
\begin{ruledtabular}
\begin{tabular}{cccccc} 
n &  1 & 2 & 3 & 5 & 10 \\ \hline
$\omega_n$ & $0.798$  & $0.926$  & $0.961$ &  $0.984$  & $0.995$  \\
$\gamma_n/q^2$ & $0.057$  & $0.010$ & $0.0035$ & $0.0009$ & $0.0001$  \\
\end{tabular}
\end{ruledtabular}
\end{table}
{\bf Collective effect of the QNM.--}
The numerical simulations in Ref.\cite{Racz} of the non-linear temporal dynamics of the
BPS monopole hit by a strong spherically symmetric pulse
have shown that a  considerable fraction of the energy received from the pulse is
not radiated away immediately. It gets `trapped' by the
monopole and forms a long living quasiperiodic excitation that radiates
very slowly, and whose late time behavior is to a large extent
independent of the structure
of the initial perturbation pulse, i.e.,\ it shows certain type of {\sl universality}.
We can now offer an explanation to these observations.

It is intuitively clear that, according to the decomposition of the eigenfunctions of the
linearized problem, there is a `radiative' sector containing the massless $h$-modes
with $\omega^2>0$ and massive $w$-modes with $\omega^2>1$, and also
a `non-radiative' sector consisting of the $w$-modes with $\omega^2<1$.
One expects that a part of the energy of the pulse received by the monopole
will be distributed among the radiative modes and will be
radiated away. However,
as a generic perturbation will have a nonzero overlap also with the non-radiative
modes, the remaining energy will get trapped in the non-radiative sector.
Since it cannot be  radiated away directly, a
long living excitation on the monopole forms that
will decay only due to a slow energy leakage to the radiative
channels.  In the terminology of black hole physics,
the perturbed monopole keeps some of its `hair' for a long time.

Although initially the dynamic of the system will be nonlinear,
one expects the nonlinear effects to become negligible
when a sufficient amount of the received energy is radiated away.
The linear approximation will therefore give a good description of the late time behavior,
providing also an explanation of the observed {\sl universality}.

Let $t=0$ be the starting point of linear regime, when
the perturbed monopole
is described by $w(0,r)=\delta W(r)$ and $h(0,r)=\delta H(r)$.
The subsequent temporal evolution is determined to a large extent
by the QNM, since they
hold their energies for a long time. Using (\ref{approx}),
we therefore approximate the general solution for $t>0$ by
\be                              \label{late}
w(t,r)=\Re\left(\sum_{n=1}^{\infty}c_n e^{-i\omega_n t-\gamma_n t}w_n(r)\right),
\ee
and similarly for $h(t,r)$. Here
$c_n=\int_0^\infty w_n(r)\,\delta W(r) dr$.
This sum of damped oscillations
corresponds to a long living excitation with the following properties.

For a localized $\delta W$ the overlap coefficients $c_n$ will be maximal
for small $n$. Therefore terms with small $n$ will dominate at first the sum in Eq.(\ref{late}).
They will soon be damped, however, since their damping rates, $\gamma_n$,
are the largest.
Terms with higher $n$ will then become more important at later times. 
This `dying out' of modes has indeed been observed in Ref.\cite{Racz} (Fig.6), 
and the values of frequencies $\omega_n$ measured there are
 in good agreement with those given in Table~I.  

\begin{figure}
\epsfxsize=9cm
\epsfysize=4cm
\centerline{\epsffile{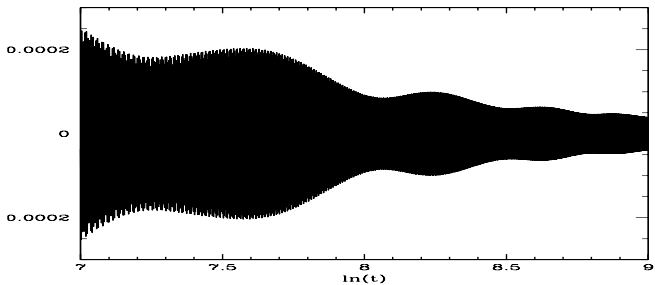}}
\caption{The profile of $w(t,1)$ in Eq.(\ref{late}) corresponding to
the initial step-like perturbation, $\delta W(r)=\theta(1-r)$.
}
\label{fig4}
\end{figure}

For any $t$
there is a number, $k(t)$, determined by the condition
$\gamma_k t\sim 1$, such that all terms with $n\ll k$ in (\ref{late}) are
already damped, while those with
$n\gg k$ are not yet important, since their $c_n$'s are small compared to $c_k$.
The sum will
therefore be dominated by terms with $n\sim k(t)$.
Considering for simplicity the sum of only two of these terms,
with frequencies $\omega_k$, $\omega_{k+1}$, gives beats with the base frequency
$\omega(t)=\frac12(\omega_{k+1}+\omega_{k})$ whose amplitude is
modulated with the frequency $\Omega(t)=\frac12(\omega_{k+1}-\omega_{k})$.
This explains qualitatively the behavior shown in Fig.\ref{fig4}.
Since $\gamma_k$ decreases with $k$, the value of $k(t)$
will grow with $t$, and so will do the base frequency $\omega(t)$.
Since $\gamma_k\sim k^{-3}$
for large $k$, it follows that  $k(t)\sim t^{1/3}$ for large $t$.
Using (\ref{llarge}), we conclude that for large $t$
one has $1-\omega^2(t)\sim t^{-2/3}$, which explains the feature observed 
in Ref.\cite{Racz} (Fig.5). 
In a similar way we obtain for the modulation frequency
$\Omega(t)\sim t^{-1}$.

Even though each term in the sum (\ref{late}) tends to zero exponentially
fast, the sum as a whole decreases considerably slower.
Since for large $t$ only terms with large $n$ are relevant in (\ref{late}),
we can use Eqs.(\ref{large}),(\ref{llarge}) to obtain
$\omega_n\approx 1-{1}/{2n^2}$  and $w_n(r)\approx n^{-3/2}w_\infty(r)$
for $r\leq \bar{r}=\int_0^\infty rw_n^2dr\sim n^2$. For a localized $\delta W$
this implies that
$c_n\approx {\cal N}n^{-3/2}$ with ${\cal N}=\int_0^\infty w_\infty\delta Wdr$.
As a result, for $t\to\infty$ and for $r<\bar{r}(t)\sim n(t)^2\sim t^{2/3}$
the solution (\ref{late}) reduces to $w(t,r)={\cal N}w_\infty(r)\Re(G(t))$, with
\be                                        \label{final}
G(t)=e^{-it}\sum_{n> t^{1/3}}^{\infty}\frac{1}{n^3}
\exp\left(\frac{it}{2n^2}-\frac{bt}{n^3} \right).
\ee
This shows that the late time dynamics is indeed {\sl universal}, since 
changing the initial conditions affects only the
normalization ${\cal N}$. 
 The final task is to determine the asymptotic behavior
of $G(t)$. Transforming the sum (\ref{final}) to a contour integral
and using the saddle point method, we find that $|G(t)|\sim t^{-5/6}$ for $t\to\infty$.
This $t^{-5/6}$ exponent
explains the feature observed in Ref.\cite{Racz} (Fig.5).

The perturbed monopole thus ends up
in a long living breathing state
dominated by the confined, slowly radiating massive modes of the gauge field.
In the region $r\leq t^{2/3}$
this state is characterized by modulated pulsations
whose base frequency approaches the vector boson mass, while the
amplitude decreases as $t^{-5/6}$.

The total energy of the breather can be obtained by summing over the QNM, 
$E=\sum_n c_n^2\omega_n^2 e^{-2\gamma_n t}
\sim
\sum_n n^{-3}\exp(-2bn^{-3}t)\sim
t^{-2/3}$.
This includes the energy of the confined massive $w$-modes and also that of
the massless $h$-radiation emitted by these modes.
$E$ decreases, since there is a  flux of the $h$-radiation at  infinity, 
$S=\dot{E}\sim t^{-5/3}$. This value agrees with the result of \cite{Racz}.
Specifically,  Fig.2 in \cite{Racz} shows
the flux $S\sim T^{-5/3}$ at future null infinity, with
$T\sim t-r$. 
By continuity, the flux
through a 2-sphere of a very large but finite radius $r$ is still $S\sim T^{-5/3}$.
But since $t$ is finite then, for $t\gg r$ one has $T\sim t$,  which agrees with  out result
$S\sim t^{-5/3}$.

In conclusion, we have found resonant excitations of the monopole
and studied some of their applications. Although we have considered above only
the BPS monopole, the existence of similar resonances can also be shown
in the case
of a non-zero Higgs self-coupling $\lambda$.

{\bf Acknowledgments.--}
We would like to thank Gyula Fodor and Istv\'an R\'acz
for communicating to us their results prior to publication and
stimulating discussions.

\end{document}